\documentclass[a4paper,11pt]{article}

\def\<<{``}	

\def\<<{°}

\usepackage{amsmath}

\usepackage[dvips]{graphicx}

\title{\textbf{Metrological characterization of the pulsed Rb clock with optical detection}}
\author{S. Micalizio, C. Calosso, A. Godone, and F. Levi \\ {\small Istituto Nazionale di Ricerca Metrologica, INRIM, Strada delle Cacce 91, 10135 Torino, Italy}}

\begin{document}

\maketitle

\section*{Abstract}

We report on the implementation and the metrological characterization of a vapor-cell Rb frequency standard working in pulsed regime. The three main parts that compose the clock, physics package, optics and electronics, are described in detail in the paper. The prototype is designed and optimized to detect the clock transition in the optical domain. Specifically, the reference atomic transition, excited with a Ramsey scheme, is detected by observing the interference pattern on a laser absorption signal.

\
The metrological analysis includes the observation and characterization of the clock signal and the measurement of frequency stability and drift. In terms of Allan deviation, the measured frequency stability results as low as $1.7\times 10^{-13} \ \tau^{-1/2}$, $\tau$ being the averaging time, and reaches the value of few units of $10^{-15}$ for $\tau=10^{4}$ s, an unprecedent achievement for a vapor cell clock. We discuss in the paper the physical effects leading to this result with particular care to laser and microwave noises transferred to the clock signal. The frequency drift, probably related to the temperature, stays below $10^{-14}$ per day, and no evidence of flicker floor is observed.

\
We also mention some possible improvements that in principle would lead to a clock stability below the $10^{-13}$ level at 1 s and to a drift of few units of $10^{-15}$ per day.

\newpage

\section*{1. Introduction}

Atomic clocks based on vapor-cell technology are receiving a great attention from basic and applied research. This is basically due to the need in many today's activities for a stable frequency reference delivered by a compact, low cost and reliable device. While the current lamp-pumped Rb clocks are showing good performances and reliability both in-orbit and on ground applications, a number of innovative clock schemes has been proposed exhibiting the capabilities to improve the stability of vapor-cell based clocks to the level of the passive hydrogen maser (PHM) or even better \cite{esnault2010, micalizio2009, bandi2011}. Also, practical realizations of such schemes are expected to impact marginally on the size, mass and power consumption budgets of currently operating Rb clocks.
\

In this paper we present the implementation and the metrological characterization of a pulsed optically pumped (POP) Rb frequency standard based on a vapor cell-microwave cavity arrangement. Our prototype exhibits a frequency stability (standard Allan deviation)  $\sigma_{y}(\tau)\leq 2\times10^{-13} \ \tau^{-1/2}$, remaining below the $10^{-14}$ up to integration times $\tau \approx 10^{5}$ s and with a relative frequency drift below $10^{-14}/$day. A proper engineering of such a clock would be then extremely interesting not only in a variety of technological applications like, e.g., radio-navigation systems, synchronization of telecommunication networks, etc., but also in the basic research as local oscillator for primary frequency standards.
\

The great potential of the POP scheme in a Rb vapor-cell with buffer gas has been demonstrated in several works, see for example \cite{micalizio2009, godone2006}. It is based on the time separation of the three phases (preparation, interrogation and detection) that usually rule the operation of an atomic clock (see Fig. 1). Specifically, an intense laser pulse initially prepares the atomic sample producing the population imbalance in the two ground state hyperfine levels of $^{87}$Rb. The atoms are then interrogated with a couple of microwave pulses resonant with the clock transition (6.834 GHz) and separated by a time $T$ (Ramsey interaction). Finally, a detection window is enabled in order to detect the atoms that have made the transition. The main feature of this technique is that the atoms make the clock transition in the dark, when the laser is off and a dramatic reduction of the light-shift occurs in comparison to the more traditional double-resonance continuous approach \cite{mcguyer2009, affolderbach2005} or also to the Coherent Population Trapping (CPT) technique \cite{shah2006, zhu2000, zhu2003, godone2004, nagel1999}. In other words, in the pulsed scheme the mutual influence of laser and microwave signals is avoided and, accordingly, the coupling between microwave and optical coherences turns out negligible. In this regime, the atoms behave most likely as a pure two-level system, as in an atomic fountain, for example. Moreover, the Ramsey interrogation technique allows to observe a clock signal with very narrow linewidth that, in addition, results nearly insensitive to any working parameter, like laser and/or microwave power, etc..
\

The detection can take place either in the microwave domain (passive maser) or in the optical domain. 
The first approach has been extensively studied in our previous works. In particular, the POP Rb maser exhibits an Allan deviation of $\sigma_{y}(\tau) = 1.2\times10^{-12} \tau^{-1/2}$ for averaging times $\tau$ up to $10^{5}$ s, and reaches the value of $5\times10^{-15}$ after drift removing at $\tau=50000$ s \cite{godone2007}, a result very close to a PHM performance. In the case of the optical detection, the laser is again on during the detection phase but it is used as a probe: duration and intensity are in general reduced from those used in the pumping phase. 
\

Qualitative physical arguments lead to consider that the detection of the clock transition through the laser absorption signal offers some advantages with respect to the maser approach. First, due to the fact that optical photons carry more energy than microwave photons, a higher signal-to-noise ratio is expected and, provided other noise sources are controlled, also a better short-term frequency stability. Moreover, the microwave cavity is used in the interrogation phase to couple the microwave signal to the atoms and does not play any role  during the detection process. This means that a high-$Q$ cavity, mandatory in the case of the POP maser, is not more required, with great benefit for the medium-long term performances since cavity-pulling turns out negligible. 
\

The system described in this paper was designed and optimized to detect the clock transition in the optical domain. The characterization here reported includes an analysis of the clock frequency dependence to the variation of the working parameters and the evaluation of the respective conversion factors. In order to reduce those sensitivities, some technical solutions, described in the paper, were adopted in the clock design. 
Particular care is devoted to the effects limiting the frequency stability performances, such as phase noise of the interrogating microwave signal and laser noise sampled during the detection process. 
\

In the paper we will often refer to the theory reported in \cite{micalizio2009}. That theory is based on a multilevel atomic model where the full Zeeman manifold of the ground and excited states is taken into account. The model also considers the dynamics induced among the Zeeman sublevels by the relaxation processes like buffer gas, spin-exchange and cell-walls collisions. 
(The theory has been developed for $D_{1}$ transition as optical pumping line, however, most of the conclusions can be extended also to $D_{2}$ line).
\

We point out that the multilevel model is needed to provide an accurate evaluation of the pumping efficiency and of the clock signal features, such as contrast, signal-to-noise ratio, etc.. The naive three-level model (two ground-state clock levels and one excited state) gives in fact only a qualitative description of the atomic sample behavior inside the cell and fails when a quantitative comparison between theory and experiment is made.
Confident of the good agreement between theory and experiments already proved in the case of the maser approach, we extended the model to the optical detection to identify the values of the working parameters, like operation temperature, laser power, microwave power, etc. that offer the best clock performances. Here we show that multilevel model is suitable to well reproduce the experimental results also in this case. 

\
The paper is organized as follows. We describe in next sections the implementation and the characterization of the three main parts of the prototype: physics package (section 2), optics (section 3) and electronics (section 4). In Section 5 we recall the principle of operation of the POP clock, discussing the optimization of some working parameters. Section 6 is devoted to the characterization of the entire apparatus working as a clock, reporting the sensitivity to the working parameters and the respective conversion factors, the drift and the frequency stability.
On the basis of the achieved results and of the theoretical predictions, in the last section we propose some possible improvements that can possibily lead to a frequency stability of $\sigma_{y}(\tau)\approx 5\times 10^{-14} \tau^{-1/2}$, down to the $10^{-16}$ region, maintaining the frequency drift of few units of $10^{-15}$ per day.

\
\begin{figure}[!h]
\begin{center}
\includegraphics[height=200pt]{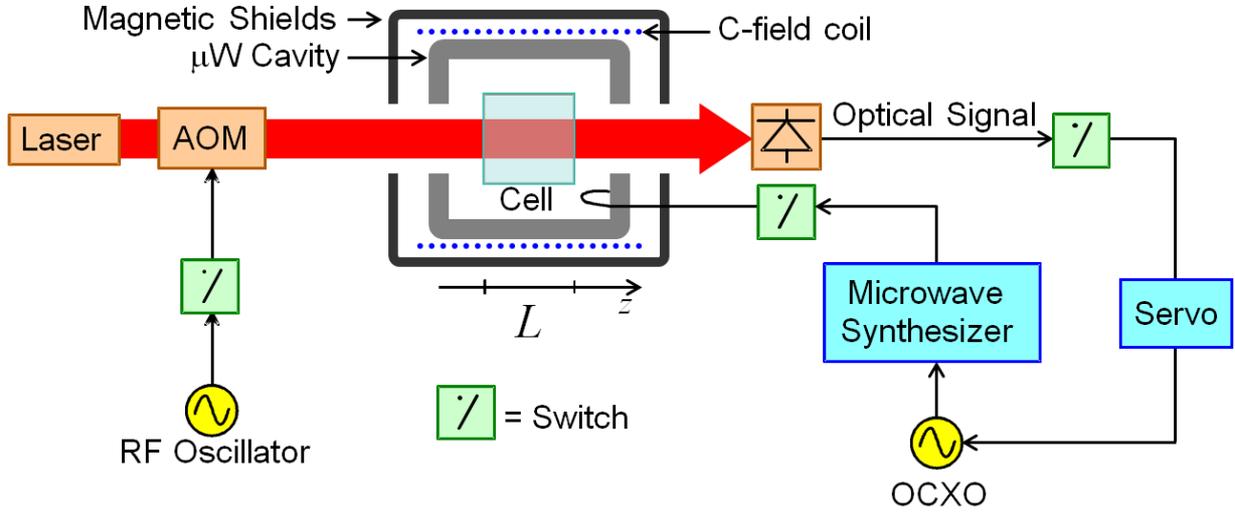}
\caption{Schematic setup of the POP clock with optical detection.}\label{}
\end{center}
\end{figure}

\section*{2. Physics package}

Figure 2 shows a schematic of the physics package. It has a layer structure and from the innermost one to the outmost the following components can be recognized:

\begin{figure}[!h]
\begin{center}
\includegraphics[height=300pt]{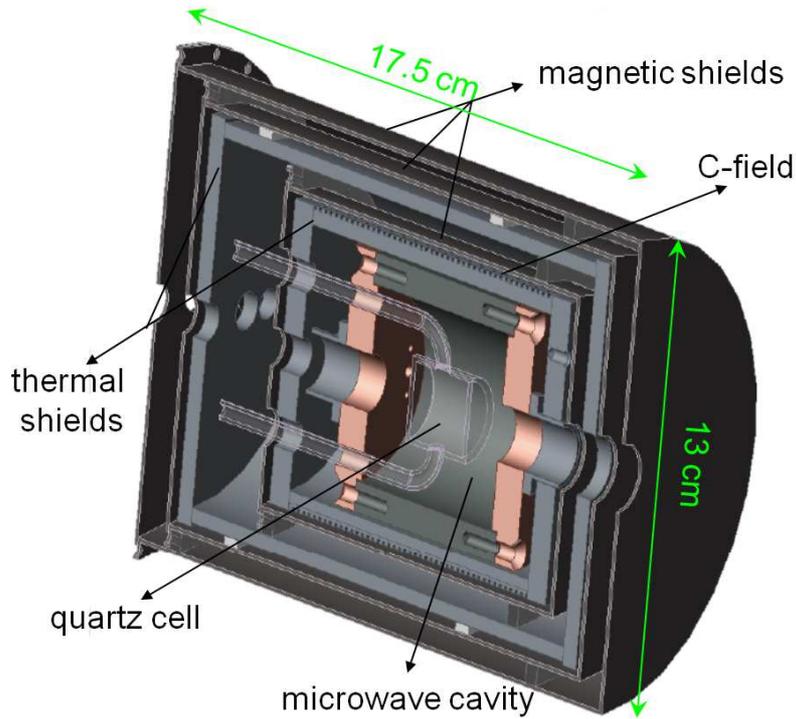}
\caption{Schematic of the physics package (longitudinal section).}\label{}
\end{center}
\end{figure}

\begin{itemize}
	\item quartz cell;
	\item microwave cavity;
	\item C-field solenoid;
	\item internal heater;
	\item first magnetic shield;
	\item external heater;
	\item two external magnetic shields.
\end{itemize}

Before describing the physics package components, we point out that the physics package has been designed in order to reduce the inhomogeneities in the active atomic medium. In fact, according to the theory \cite{micalizio2009}, the spatial inhomogeneity makes the resonance frequencies of the Rb atoms depending on their position in the cell. The observed clock frequency is then an average of such frequencies and, in addition, it changes with laser intensity simulating an off-resonant light-shift. In this context, the spatial distribution of the cavity electromagnetic mode plays a fundamental role since it introduces a coupling, otherwise absent, between the power of the externally applied microwave and the power of the pumping laser. To reduce this phenomenon known as positioning-shift \cite{english1978, alekseyev1974, alekseyev1975}, the microwave field must be uniform in the region where the atomic sample is placed. 
\

Of course, the uniformity in the cell is also required from a thermal point of view.

\subsection*{2.1 Cell}

The core of the clock is a quartz (fused silica) cell with a radius $R= 10$ mm and a length $L= 20$ mm. The cell has two filling stems placed in symmetric position with respect to its sagittal plane.  This allows for a precise positioning of the cell in the center of the cavity, avoids axial rotation and improves the symmetry of the system. 
\

The cell is filled with $^{87}$Rb atoms and a mixture of buffer gases, Ar and N$_{2}$ in the pressure ratio 1.6, with a total pressure $P_{t}$ of 25 Torr (3.3 kPa). The mixture is nominally temperature-compensated at $65 \ ^\textnormal{0}$C.  According to theory \cite{micalizio2009}, higher temperatures would significantly increase the laser absorption through the cell along its propagation direction ($z$ axis) with a consequent deterioration of the uniformity inside the cell. Moreover, in high-density media the central Ramsey fringe may suffer of distortion effects which introduce a dependence of its linewidth on the atomic density and on the laser intensity.
Taking into account the cell size, the buffer gas composition and the working temperature ($66 \ ^\textnormal{o}$C, see Sec. 2.4), it is possible to estimate the relaxation rates of atomic population and coherence in the ground state $\gamma_{1}= 360 \ \textnormal{s}^{-1}$, $\gamma_{2}= 300 \ \textnormal{s}^{-1}$, respectively. The relaxation rate of the excited state turns out $\Gamma^{*}= 3 \times 10^{9} \  \textnormal{s}^{-1}$ and the cell optical length $\zeta=17$ for $D_{2}$ optical transition. 
\

As is well known, the buffer gas induces a shift in the clock transition \cite{vanier_audoin1985, vanier1982} that in our case is about $171$ Hz/Torr. The expected shift is then of the order of $4.3$ kHz and turns out in good agreement with the measured value.

\subsection*{2.2 Cavity}

%We recall that the cavity has been designed in order to resonate when empty at a frequency higher than the atomic frequency. In particular, in our case the cavity resonates at  C =7.659 GHz when empty. The effect of the quartz %cell is to shift the resonance frequency of  860 MHz.
%To exactly tune the cavity on the resonance frequency, three tuning steps are required:

%The gap in the cavity resonance frequency is   48.8 kHz/°C and this value accounts not only for the thermal expansion of the cavity but also for the variation of the cell thermorefractive index 

The microwave cavity has a cylindrical geometry with internal diameter $2 a= 52$ mm and internal length $d= 49$ mm. It is made of molybdenum and resonates on the electromagnetic mode TE$_{011}$ at the $^{87}$Rb ground-state hyperfine frequency (6.834 GHz). The cavity is designed with mode chokes in the end-caps to suppress the degenerate electromagnetic mode TM$_{111}$.
The relative size of the cell-cavity arrangement was optimized and comes from a trade-off between two opposite requirements: a) the atomic sample should fill a small portion of space in the center of the cavity to assure a good uniformity of the microwave field experienced by the atoms; b) a  cell large enough is required to assure a large number of interacting atoms. The aspect ratios $2a/d \approx 2R/L \approx 1$ satisfy both these requirements.
\

%The choice of the cavity material is motivated by the need to guarantee a good frequency stability also in the medium-long term: although heavier than other metals like Al, for example, Mo has in fact thermal expansion coefficient %that is four times lower. Moreover, drift effects related to aging of the material (dislocations in the crystal and creeping) are significantly lower due to the higher melting temperature \cite{ferro_milone1980}. This allows for %reduced cavity tuning requirements and longer operation times of the clock. 

The cavity loaded quality factor is $Q_{L}\leq 1000$ and is mainly defined by the condensation of Rb films on the inner surface of the cell. The cavity coupling parameter is $\beta \approx  0.05$.

%The current loaded quality factor of the cavity $Q_{L}$ is below 1000 and the cavity coupling parameter is 
%We point out that the cell is in quartz to reduce the dielectric losses. As previously mentioned, a low $Q_{L}$ does not compromise the clock operation, however, as mentioned before, even in the case of optical detection it is of %crucial importance to maintain a well defined cavity mode in order to achieve the maximum homogeneity in the field parameters. 
%We think that the current value of $Q_{L}$ ($<1000$) is limited by the condensation of Rb films on the inner surface of the cell. Indeed, to avoid this effect, a cold point in the cell has been established; the two filling stems %are in fact in direct contact with a thermal shield that is maintained at a lower temperature with respect to the cell body (see later). However, despite this temperature gradient in the cell, the cavity $Q_{L}$ is continuously %decreasing. We attribute this behavior to the presence of impurities and/or manufacturing defects in the cell that may act as trapping centers for Rb atoms, allowing the atomic deposition. As a consequence, also the cavity coupling %parameter turns out very low $\beta \approx 0.05$. 
\
\begin{figure}[!h]
\begin{center}
\includegraphics[height=250pt]{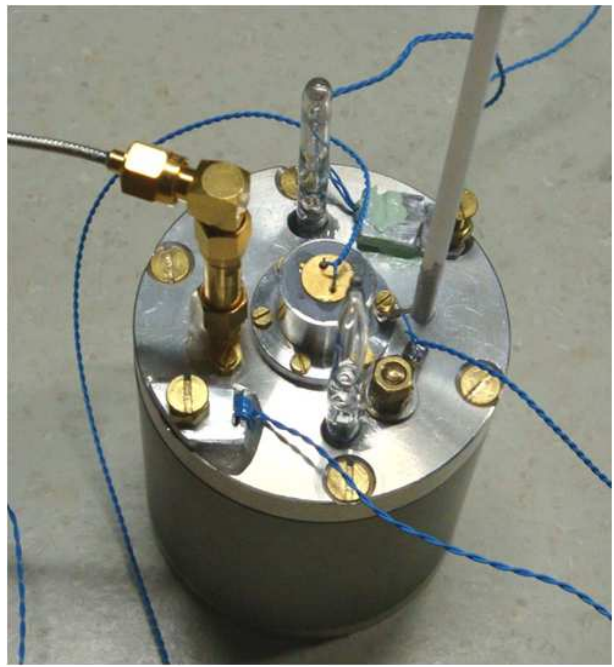}
\caption{The cell-cavity system. It is possible to recognize the two stems of the cell, the photodiode, the tuning screw, the microwave cable and the NTC's sensors.}\label{}
\end{center}
\end{figure}

The cell-cavity arrangement (see Fig. 3) here employed has been characterized in more detail in \cite{godone2011} (it corresponds to the cavity system called $C_{2}$ in that paper). That cavity is equipped with both a PIN diode (used for $Q$-switching technique) and a GaAs varactor diode. In the experiments reported in this paper we used only the latter for electronic tuning of the cavity. 

\subsection*{2.3 Quantization magnetic field}

The cavity is placed inside an Al cylinder that holds a solenoid generating a C-field aligned with the cavity axis ($\hat{z}-$axis). The quantization magnetic field $B_{0}$ is around $1.5 \ \mu$T. This value assures a negligible Rabi pulling frequency shift due to $\pi-$transitions with $m_{F}\ \neq 0$ and avoids the generation of Zeeman coherences in the two ground-state manifolds.

\subsection*{2.4 Heaters and thermal sensitivity}

Temperature fluctuations can be transferred to the clock transition mainly through the buffer gas; specifically, for a thermally uniform cell and for our temperature-compensated buffer gas, the expected relative frequency sensitivity is $(\Delta \nu/\nu)/\Delta T \leq 1 \times10^{-11}/^\textnormal{o}$C around $T= 65 \ ^\textnormal{o}$C \cite{micalizio2010}. However, dynamical temperature inhomogeneities in the cell due to the stems worsen the above sensitivity to the measured value of $(\Delta \nu/\nu)/\Delta T \approx 1 \times10^{-10}/ ^\textnormal{o}$C. 

\begin{figure}[!h]
\begin{center}
\includegraphics[height=250pt]{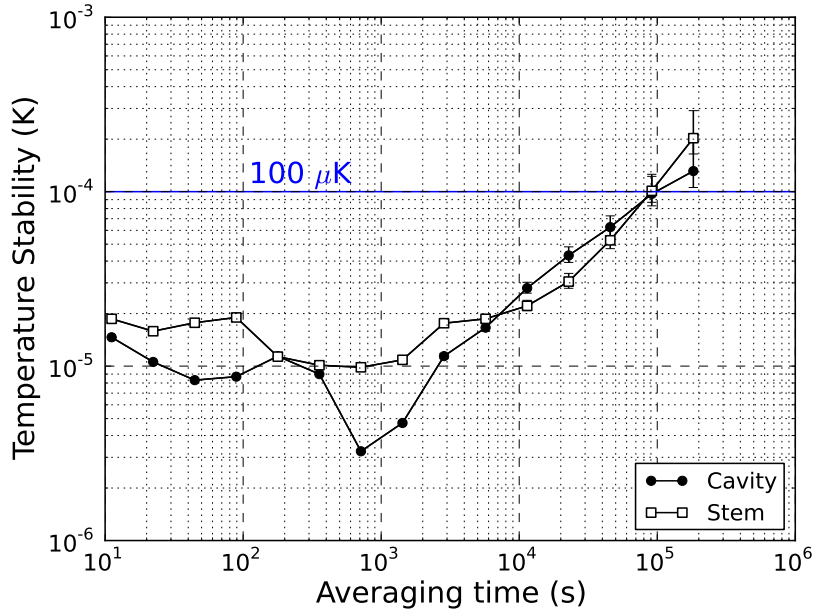}
\caption{Temperature stability of the internal and of the external thermal shields as measured by two NTC used as monitor.}\label{}
\end{center}
\end{figure}

\
This sensitivity implies that the temperature should be stabilized at a level of $100 \ \mu$K in order to reach a frequency stability in the $10^{-14}$ range. To keep the temperature stable at that level maintaining at the same time the required gradient between the stems and the cell body, two active controls have been implemented.

\
The first temperature-controlled element is the same Al cylinder (innermost thermal shield) used to sustain the C-field solenoid; a heater is in fact wrapped on it. The heater is AC-driven ($\approx$ 40 kHz) to avoid the generation of spurious magnetic fields, and works at a temperature of $T_{int} = 68 \ ^\textnormal{o}$C during clock operation. 
\

The cylinder is placed inside a mu-metal magnetic shield which in turn is housed in a second Al cylinder supporting the external heater. This second heater may be DC- or AC-driven and its operation temperature is $T_{ext} = 66 \ ^\textnormal{o}$C. 
\

Figure 4 shows that both the temperatures are stable at the level of $100 \ \mu$K for integration times up to one day.

\subsection*{2.5 Magnetic shields and vacuum enclosure}

The physics package is completed by two magnetic shields; each magnetic shield rejects the external longitudinal magnetic fluctuations by a factor of 10, so the overall shielding factor is around 1000 along the $z-$axis. In quiet geomagnetic conditions, this shielding factor is sufficient to guarantee stable magnetic field  at the level of 10 pT. The shields magnetic noise is negligible in our setup.

\
The physics package is housed in a vacuum enclosure to isolate the apparatus from the environmental fluctuations.  As explained in \cite{micalizio2010}, the environment pressure fluctuation is responsible of a change in the refractive index of the air: 1 Pa gives rise to a relative variation of the cavity resonance frequency of $-2.6\times10^{-9}$ (17 Hz/Pa). Humidity in the air is also responsible of the change of the resonance frequency \cite{godone2011}. The vacuum is maintained below $10^{-3}$Pa; this value guaranties also the real operation of the physics package under vacuum conditions from a thermal point of view.
\

\section*{3. Optics}

We made experiments with lasers exciting both the $D_{1}$  (795 nm) and the $D_{2}$ (780 nm) optical lines \cite{micalizioFCSEFTF}. In this paper we report only the data referring to the set-up operating at 780 nm that offers the best clock performances.
\

The laser source is a DFB diode with a linewidth of about 20 MHz limited by the current noise of the power supply. It is frequency stabilized through a third harmonic frequency control loop to the crossover transition $F=1\rightarrow F'=1, 2$ via a saturated-absorption signal. The locking dip is observed in a reference cell containing isotopically enriched $^{87}$Rb. The laser is linearly polarized and propagates along the cell $z$-axis, therefore, due to selection rules, the atoms see the light as a combination of $\sigma^{+}$ and $\sigma^{-}$ polarizations. The main part of the laser beam is sent to the physics package through an acousto-optic modulator (AOM) operating in double pass configuration at an RF frequency of 85 MHz. In this way, the laser frequency matches the absorption frequency of the atoms in the cell that results red-shifted by the buffer gas pressure (this shift is  $\approx-6.8$ MHz/Torr and in our cell is about $-170$ MHz). Moreover, the AOM also acts as an optical switch for the pulsed operation. The laser power extinction ratio $P_{L}^{on}/P_{L}^{off}$ is 45 dB. 

\
The laser delivers up to 15 mW at the entrance of the cell; the laser beam has a diameter $\phi \approx 15$ mm. An a-spheric lens with a diameter of 11 mm and a focus of 8 mm focuses the laser transmitted through the cell onto the detection photodiode placed outside the cavity.

\
In the following, in order to compare the experimental results with the theoretical predictions, we will express the laser intensity in terms of the pumping rate defined as $\Gamma_{p}=\omega_{R}^{2}/2 \Gamma^{*}$, where $\omega_{R}$ is the (angular) optical Rabi frequency. The pumping rate is related to the laser intensity $I_{L}$ through the relation:

\begin{equation}
\Gamma_{p}=\frac{Z_{0}}{\Gamma^{*}}\left(\frac{d_{e}}{\hbar}\right)^2 I_{L}
\end{equation}
\

where $d_{e}$ is the electric dipole moment of the optical transition, $Z_{0}$ is the impedance of free space, and $\hbar$ the reduced Planck constant. From Eq. (1), it results that an intensity of $1 \ \textnormal{mW}/\textnormal{cm}^{2}$ corresponds to a pumping rate $\Gamma_{p} \approx 10^{5} \ \textnormal{s}^{-1}$ for $D_{2}$ transition.

\
We also measured the laser relative intensity noise (RIN) of the laser; its noise spectral density is:

\begin{equation}
S_{a}(f)=\left(8\times 10^{-12}f^{-1}+3\times 10^{-15} \right)\ \textnormal{Hz}^{-1}
\end{equation}

$f$ being the Fourier frequency expressed in Hz. This measurement will be discussed later on in the section devoted to the clock performances analysis.

\section*{4. Electronics}

Figure 5 shows the clock scheme from an electronic point of view. The architecture is similar to the one we developed for the POP maser \cite{calosso2007}, with some important differences. In the optical detection the laser is not only used to pump the atoms but also to detect the clock resonance; the atomic signal provided by the photodetector is processed by a trans-impedance amplifier followed by a signal conditioning circuit to match the input signal dynamics of the lock-in amplifier. The synthesis chain was redesigned in order to support the expected stability. In particular, we require that the phase noise associated to the microwave signal should not contribute to the clock stability above the $1\times 10^{-13} \tau^{-1/2}$ level.

\begin{figure}[!h]
\begin{center}
\includegraphics[height=250pt]{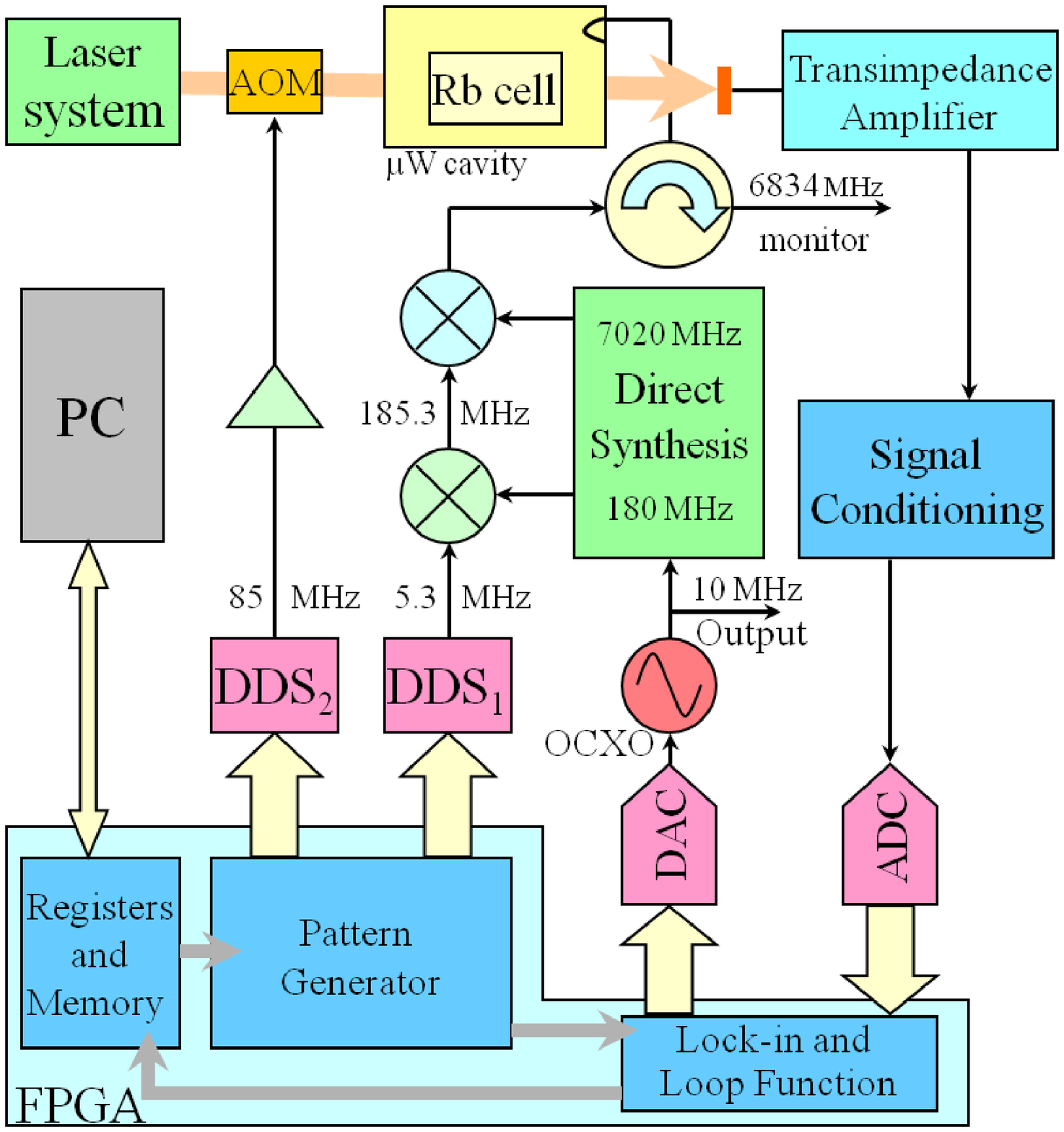}
\caption{Electronic architecture of the POP clock.}\label{}
\end{center}
\end{figure}

\subsection*{4.1 Synthesis chain}

The synthesis chain provides the two phase-coherent microwave pulses at 6.834 GHz required by the Ramsey interaction scheme, starting from an OCXO at 10 MHz operating as local oscillator (LO). 
The signal at 6.8347 GHz is generated by synthesis and mixing of three signals at 5.3 MHz, 180 MHz and 7020 MHz.
\

From the phase noise point of view, the most critical part is related to the first multiplication stages composed of a X9 direct frequency multiplier from 10 MHz to 90 MHz, followed by a X2 direct frequency multiplier from 90 MHz to 180 MHz. Both the multipliers are based on Schottky diodes to comply with the low phase noise requirement. A PLL at 180 MHz acts as a filter for the unwanted spurs. The signal at 180 MHz drives a non-linear transmission line (NLTL) that works as a comb generator. In comparison to the solution adopted in \cite{calosso2007} employing a step recovery diode (SRD), the NLTL is nominally 20 dB less noisy \cite{boudot2009}. In particular, for our device, the white phase-noise level is $-178 \ \textnormal{dBrad}^{2}/\textnormal{Hz}$ at 200 Hz (referred at 10 MHz). 
The 39-th harmonic of the comb is selected by a coaxial filter and is used as reference for the PLL that locks a YIG oscillator at 7020 MHz. 
The 5.3 MHz signal is synthesized by a DDS (indicated as DDS1 in the figure) and is mixed to the 180 MHz signal to obtain 185.3 MHz that is, in turn, subtracted to the 7020 GHz output to match the $^{87}$Rb frequency at 6.834 GHz. 

\begin{figure}[!h]
\begin{center}
\includegraphics[height=250pt]{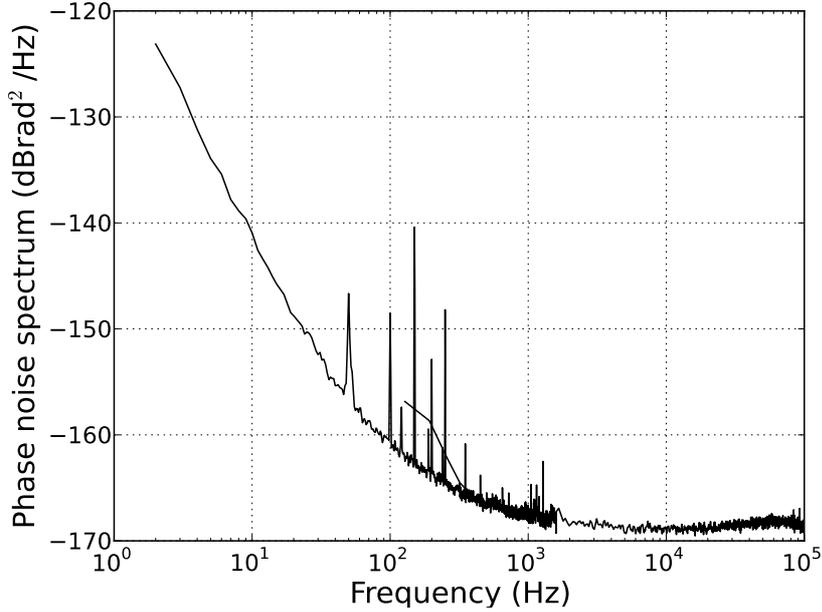}
\caption{Phase noise power spectrum of the synthesis chain and OCXO referred to a 10 MHz carrier.}\label{}
\end{center}
\end{figure}

The phase noise power spectral density $S_{\phi}(f)$ was measured versus a second identical chain with the homodyne technique and, referred to a 10 MHz carrier, turns out:

\begin{equation}
S_{\phi}(f)=\left(1.26\times10^{-17}+4\times10^{-15} f^{-1}+3.1\times10^{-13}f^{-2}+3.1\times10^{-12}f^{-3}\right) \ \textnormal{rad}^2/\textnormal{Hz}
\end{equation}

where $f$ is the Fourier frequency. The $S_{\phi}(f)$ behavior is reported in Fig. 6.

\subsection*{4.2 Trans-impedance amplifier}

A 15 mm$^2$ silicon photodiode followed by a transimpedance amplifier of gain 1.6 V/mA is used to detect the light transmitted through the cell ($\approx 1$ mW) during the detection stage. It is able to operate with power levels up to 10 mW avoiding, at the same time, a degradation of the clock stability. The noise performance of this circuit is only limited by the thermal noise of the feedback resistor thanks to the use of an operational amplifier based on ultra low noise BJT technology.
%has been designed in order not to destroy the cavity quality factor. In fact, even in the case of optical detection it is of crucial importance to maintain a well defined cavity mode in order to achieve the maximum homogeneity in %the field parameters. Any inhomogeneity will result in an increased clock sensitivity to external fluctuation as laser and microwave power noise conversion, temperature sensitivity etc.

\subsection*{4.3 Digital board}

The pulsed operation is performed by a dedicated digital board where a single FPGA coordinates the operation of the two DDSs and of the ADC: 1) DDS1 generates the baseband version of the two microwave pulses of the Ramsey scheme (see Figure 5); DDS2 drives the AOM during the pumping and the detection phases, 3) ADC is the front-end of the lock-in amplifier.
The frequency loop controller of the LO is also implemented in the FPGA. The working parameters are set by a PC that is used to monitor the operation, but it is not a part of the running clock.

\section*{5. Operation and timing}

Before reporting the results obtained with the system described above, we recall the principle of operation of the POP clock.

\
As mentioned in the introduction, the POP clock operation separates in the time the three following phases: 1) laser optical pumping, 2) microwave interrogation and 3) detection of the clock transition. Figure 7 shows a typical timing sequence made explicit in the case of the optical detection; $t_{p}$ is the pumping time, $t_{1}$ is the duration of each microwave pulse (Rabi time), $T$ represents the time interval where the atoms are allowed to evolve freely (Ramsey time) and $t_{d}$ is the detection time in which the clock transition is observed through the laser absorption signal.
\

\begin{figure}[!h]
\begin{center}
\includegraphics[height=250pt]{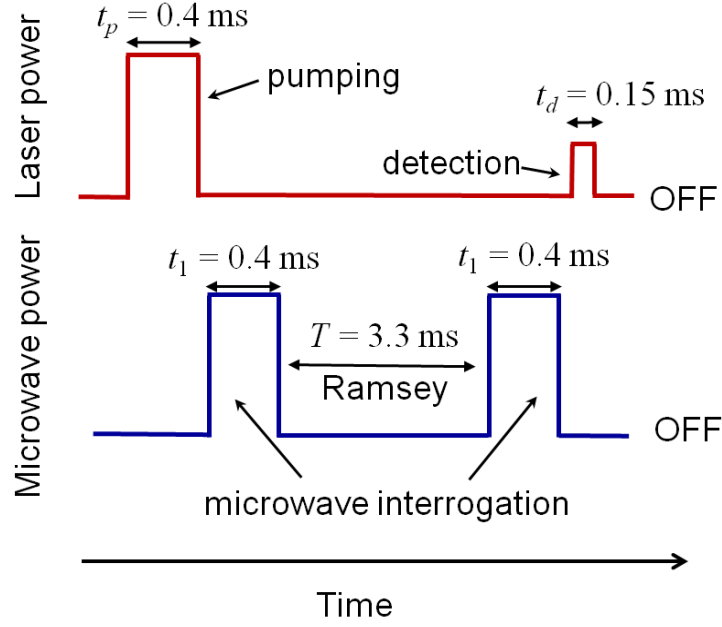}
\caption{Timing sequence in the POP clock with optical detection of the clock transition. Numerical values refer to the operating conditions of the experiments reported in the next section. (Times are not in scale).}\label{}
\end{center}
\end{figure}

The optimization of the timing sequence in the POP scheme is not easy from a mathematical point of view since many parameters (laser power, atomic density, microwave field uniformity, etc.) contribute to the clock signal, and often they appear entangled among them. However, some basic requirements suggested by the theory reported in \cite{micalizio2009} provide the guidelines to optimize the clock signal and then the stability performances.

\
First, the laser pulse during the pumping phase should produce a large population inversion among the two clock levels and, at the same time, should destroy effectively the microwave coherence to avoid residual light-shift; both the conditions are satisfied when $\Gamma_{p}t_{p}\gg 1$ \cite{micalizio2009}.

\
As regards the Ramsey pulses, the microwave power is adjusted in order to maximize the contrast of the central Ramsey fringe; in terms of pulse area this corresponds to $b_{e}t_{1}\approx \pi/2$, where $b_{e}$ is the angular Rabi frequency associated to the microwave field.

\
The Ramsey time is basically limited by the relaxation phenomena taking place inside the cell; in particular, a Ramsey time $T$ of the order of $\gamma_{2}^{-1}$ is a good trade-off between a large clock signal and a narrow fringe.

\
In the detection phase the laser pulse should minimally perturb the atoms, being mainly used as a probe: intensity and duration are reduced consequently. 

\
Provided the previous points are satisfied, $t_{p}$, $t_{1}$ and $t_{d}$ should be as short as possible in order to reduce the Dick effect (see Eq. (9)) that is inversely proportional to the duty cycle $T/T_{C}$, $T_{C}$ being the cycle duration: $T_{C}= t_{p}+2t_{1}+T+t_{d}$.

\
Once the previous requirements are fulfilled, the values of the working parameters are finely adjusted to minimize the frequency instability of the clock.

\section*{6. Results}

\subsection*{6.1 Ramsey fringes and contrast}

Figure 8 shows Ramsey fringes as detected by the photodiode at the exit of the cell. The laser power during the pumping period is $P_{L}^{pump}=10$ mW while during the detection time is $P_{L}^{det}=1.5$ mW. The figure refers to an internal temperature of $T_{int}= 63.5 \ ^\textnormal{o}$C and an external temperature of $T_{ext}= 54 \ ^\textnormal{o}$C. Although the optimum clock operation temperature is higher, we observed the Ramsey fringes also in the low temperature regime in order to verify the behavior predicted by the theory. The other main working parameters are indicated in the caption of the figure. In these operating conditions, the central fringe of the Ramsey pattern exhibits a contrast of 15\% and a linewidth, expressed in terms of full-width at half maximum (FWHM), of $\Delta \nu_{1/2}= 120$ Hz, according to the well known relation $\Delta \nu_{1/2}=1/2T$.

\begin{figure}[!h]
\begin{center}
\includegraphics[height=250pt]{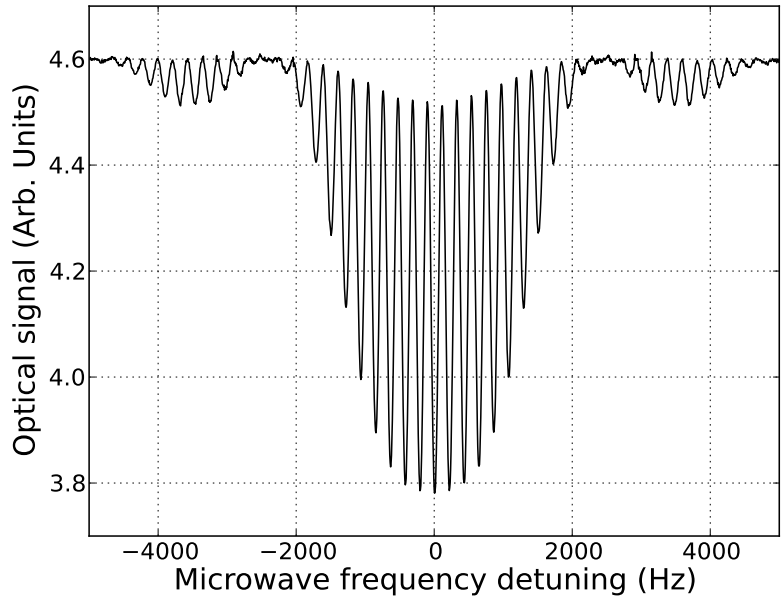}
\caption{Ramsey fringes as observed in the optical domain in the low temperature regime; $t_{p}= 4$ ms, $t_{1}= 0.4$ ms, $T=4.2$ ms, and $t_{d}=0.15$ ms.}\label{}
\end{center}
\end{figure}

\
The best clock performances were obtained at a higher working temperature ($T_{int}= 68.5 \ ^\textnormal{o}$C and $T_{int}= 66.5 \ ^\textnormal{o}$C); the corresponding Ramsey fringes are reported in Figure 9. In this case we used $P_{L}^{pump}= 15$ mW and $P_{L}^{det}=1$ mW, corresponding approximately to pumping rates of $8.5\times10^{5} \ \textnormal{s}^{-1}$ and $5.6\times10^{4} \ \textnormal{s}^{-1}$ respectively. 

\
The different Rabi envelopes of the curves reported in Figs. 8 and 9 are predicted by the theory and can be well explained in terms of the values assumed by relaxation rates $\gamma_{1}$ and $\gamma_{2}$ in the two temperature regimes, as already observed in the case of the microwave detection \cite{micalizio2009}.

\begin{figure}[!h]
\begin{center}
\includegraphics[height=250pt]{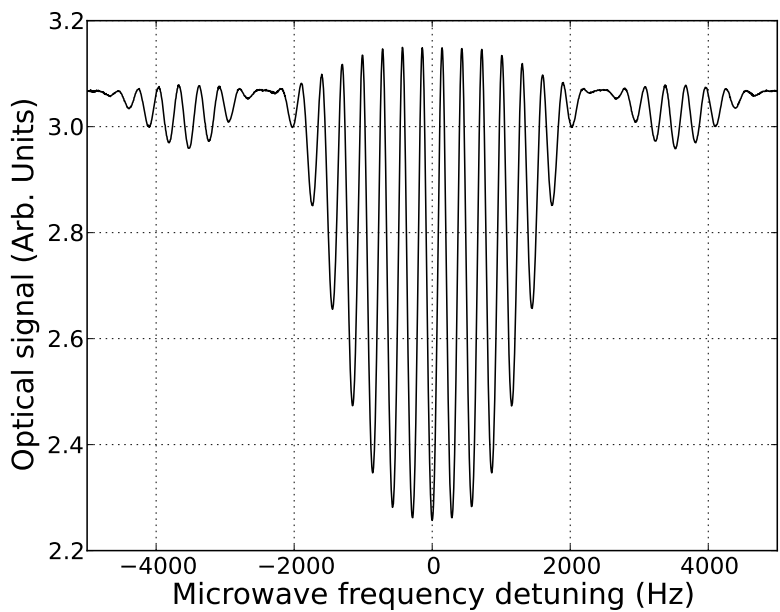}
\caption{Same as Fig. 8 but in the high temperature regime; $t_{p}= 0.44$ ms, $t_{1}= 0.4$ ms, $T=3.3$ ms, and $t_{d}=0.15$ ms.}\label{}
\end{center}
\end{figure}
\
We point out that the atomic density and the other cell parameters (relaxation rates, etc.) appear to be tightly connected to the external temperature rather than to the internal one. This can be explained taking into account that the system is under vacuum: thermal convection between the cavity and the cell is absent while irradiation and conduction through the small cell-cavity contacts are negligible. The thermal behavior is in fact driven by the diffusion motion of the buffer gas particles toward the stems and backward to the cell body; in this way, the atoms thermalize to a temperature closer to that of the external thermal shield. 

\
In the configuration of Fig. 9 the cycle time is $T_{C}= 4.65$ ms and the duty cycle is $T/T_{C} \approx 0.7$, a condition particularly favourable to reduce Dick effect (see Eq. (9)).

\
The central fringe of the Ramsey pattern shows a contrast of 28\% and a linewidth of $\approx 150$ Hz. 

\begin{figure}[!h]
\begin{center}
\includegraphics[height=450pt]{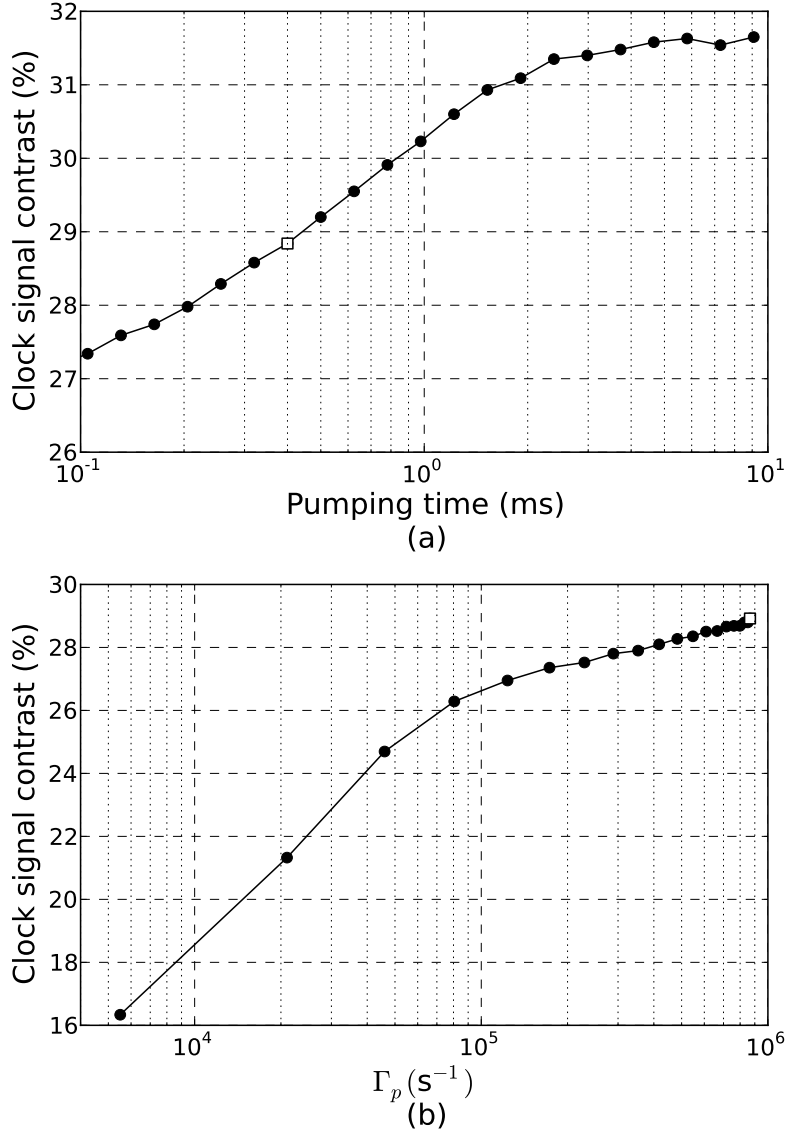}
\caption{(a) Contrast of the Ramsey fringe versus the pumping time $t_{p}$ at a fixed pumping rate $\Gamma_{p}=8.5 \times 10^{5} \ \textnormal{s}^{-1}$; (b) Contrast of the Ramsey fringe versus the pumping rate $\Gamma_{p}$  at a fixed $t_{p}=0.4$ ms.}\label{}
\end{center}
\end{figure}

\
The contrast of the central fringe was characterized in detail in the high temperature regime. In fact, it plays a key role not only to determine the signal level but also to reduce some noise sources, as will be shown later (see Eqs.(8) and (10)).

Figure 10 shows the contrast versus the pumping parameters ($t_{p}$ and $\Gamma_{p}$).

\
It is observed that, as expected, the pumping process is initially characterized by an exponential trend, after which a saturation value is reached. This level is tightly related to the maximum population imbalance which is generated in the ground state during the pumping process. In Fig. 10 and in the following ones the white square corresponds to the working point when the system operates as a clock.

\
Figure 11 shows the contrast behavior when measured versus the detection parameters (we call $\Gamma_{p}^{det}$ the laser pumping rate during the detection period). The contrast behavior can be explained taking into account that increasing the detection time and/or the detection laser power, the laser light does not act simply as a probe but gives rise to a pumping process in the atomic sample with a consequent degradation of the clock signal.

\
The contrast was also measured versus the microwave power. The maximum of Fig. 12 corresponds to $\pi/2$ pulses. Definitely, from previous figures it turns out that around typical operating conditions the observed contrast is 28-30\%; we will discuss later (subsection 6.3) how the contrast affects the clock frequency stability.

\newpage
\begin{figure}[!h]
\begin{center}
\includegraphics[height=450pt]{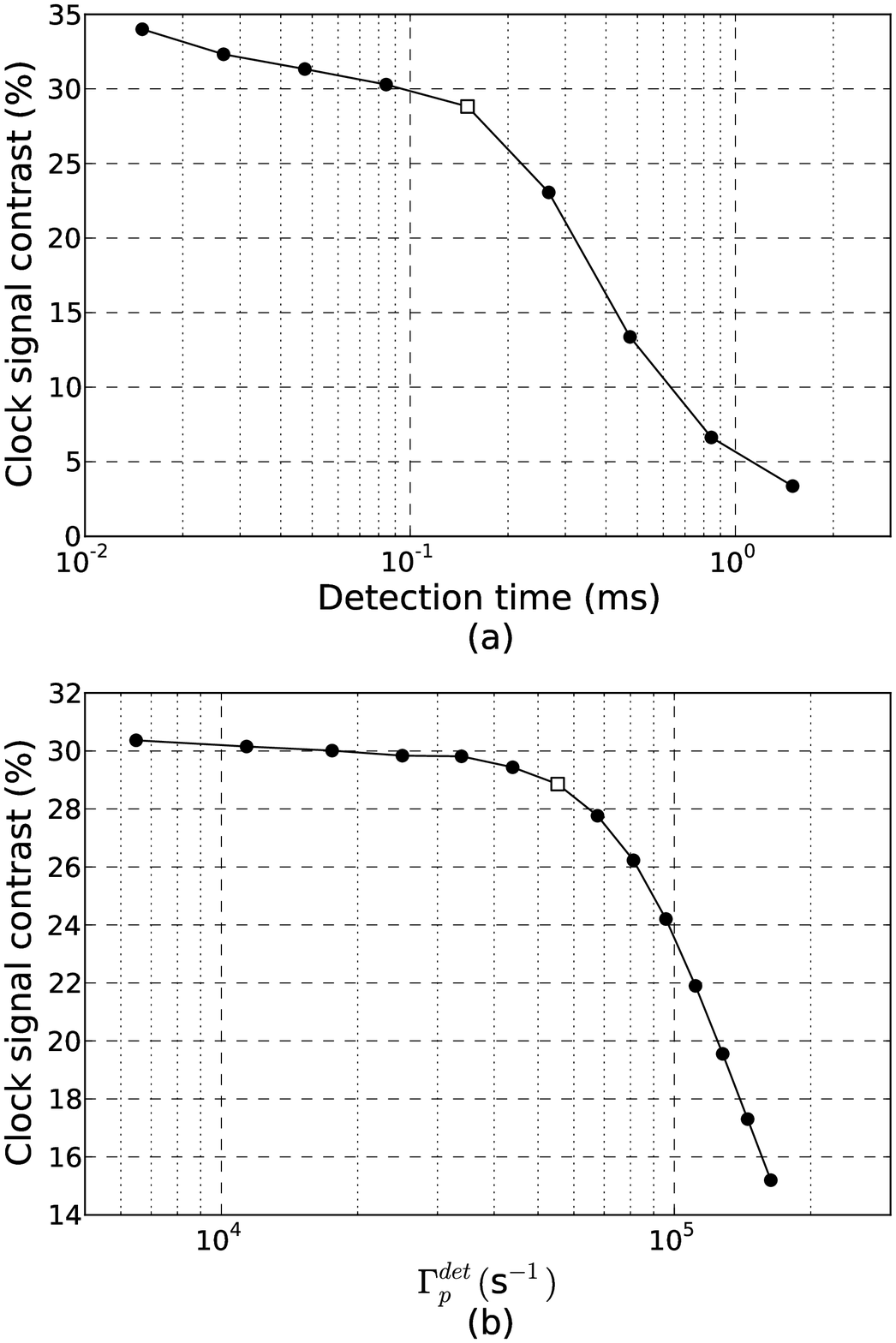}
\caption{(a) Contrast of the Ramsey fringe versus $t_{d}$ at a fixed laser pumping rate $\Gamma_{p}^{det} = 56000 \ \textnormal{s}^{-1}$; (b) contrast of the Ramsey fringe $\Gamma_{p}^{det}$ at a fixed detection time $t_{d}= 150 \ \mu$s.}\label{}
\end{center}
\end{figure}

\newpage
\begin{figure}[!h]
\begin{center}
\includegraphics[height=250pt]{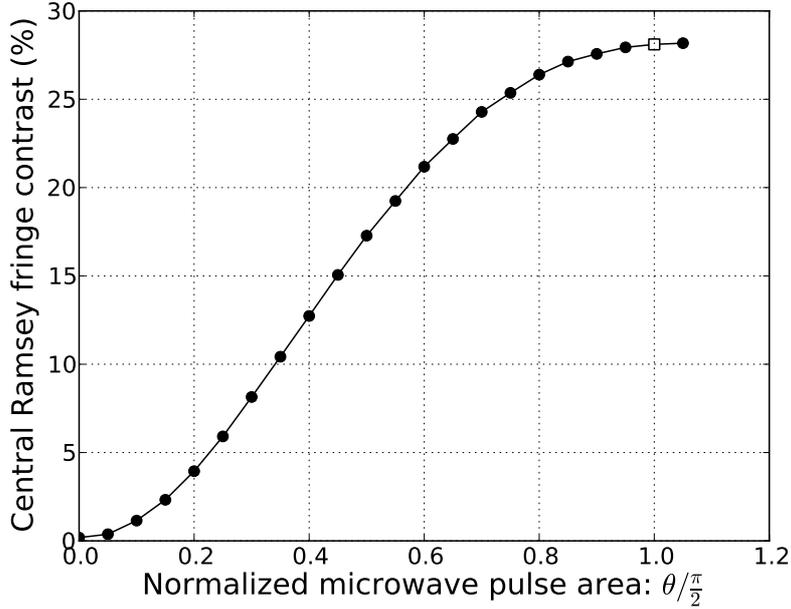}
\caption{Contrast of the clock signal versus the microwave pulse area.}\label{}
\end{center}
\end{figure}

\subsection*{6.2 Residual laser and microwave sensitivities}

The pulsed approach allows in principle to uncouple the laser pumping phase from the successive microwave interrogation, in such a way that the atomic response is insensitive to any laser fluctuations. However, due to the inhomogeneity in the atomic sample a residual coupling between optical and microwave signals exists and the clock frequency still exhibits a dependence on the laser power. Specifically, we measured the following sensitivity of the clock frequency to the laser power fluctuations $\Delta P_{L}$:

\begin{equation}
\frac{\Delta \nu /\nu}{\Delta P_{L}/P_{L}}\approx -6 \times10^{-14}/\%
\end{equation}

\
Although it is not possible to exclude totally a residual off-resonant light shift contribution, we mainly attribute this effect to a position shift that makes the resonant frequency of Rb atoms dependent on their spatial location within the cell.

\
We measured also the clock frequency sensitivity to the laser frequency fluctuations $\Delta \nu_{L}$ and we observed a residual resonant light shift of:

\begin{equation}
\frac{\Delta \nu /\nu}{\Delta \nu_{L}}\approx 1.5 \times10^{-14}/\textnormal{MHz}
\end{equation}

Anyway, it is worth mentioning that the sensitivities to the laser parameters are orders of magnitude lower than those observed in continuous operating vapor-cell clocks.

\
Although in residual form, the clock frequency suffers of a cavity-pulling shift. We recall that in pulsed operation, in the limit of small cavity detuning, the cavity-pulling shift may be written as:

\begin{equation}
\Delta\nu_{cp}=\frac{4}{\pi} \frac{Q_{L}}{Q_{a}}c(\theta)\Delta\nu_{C}
\end{equation}

where $Q_{a}$ is the atomic quality factor, $\Delta\nu_{C}$ is the cavity detuning and $c(\theta)$ is a function of the microwave pulse area $\theta = b_{e} t_{1}$ that is minimized when $\theta \approx \pi/2$ \cite{micalizio2010}. We measured a relative clock frequency change of $-2\times10^{-13}$ when $\Delta\nu_{C}=300$ kHz; for our cavity in Mo this corresponds to $\Delta \nu_{cp}/\Delta T_{int}=3\times 10^{-14}/^\textnormal{o}$C, a negligible value in our case.

\begin{figure}[!h]
\begin{center}
\includegraphics[height=250pt]{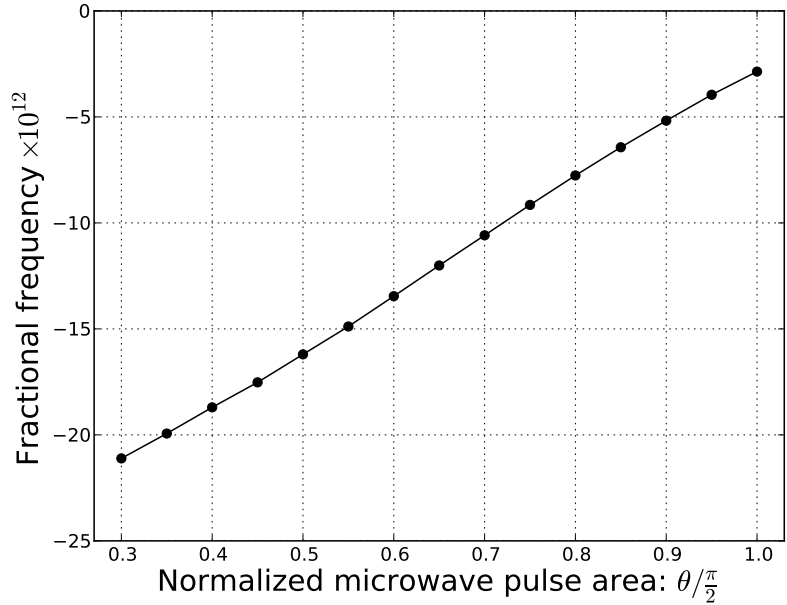}
\caption{Cavity pulling measurement: relative clock frequency versus the microwave pulse area.}\label{}
\end{center}
\end{figure}

\
Actually, Eq. (6) suggests that the cavity-pulling has another component related to the microwave pulse area. Figure 13 shows the relative clock frequency versus $\theta$; the measurement was done by changing the microwave power at a fixed value of $t_{1}$ ($t_{1}= 400 \ \mu$s).

From Fig. 13, the sensitivity of the clock frequency to the variation of $\theta$ may be expressed as:

\begin{equation}
\frac{\Delta\nu_{cp}/\nu}{\Delta\theta/\theta}\approx 5.5\times 10^{-14}/\%
\end{equation}

This sensitivity may impact the medium-long term frequency stability and an active system to stabilize the microwave power has been then implemented.

\subsection*{6.3 Frequency stability performances}

Figure 14 shows the clock frequency stability when the local oscillator (LO) is locked on the central fringe of the Ramsey pattern of Fig. 9. The figure reports also the stability of the reference frequency standard (a H-maser filtered by a BVA quartz). The figure refers to a sample of $10^6$ data, corresponding to about 12 days of measurement. At 1 s, the Allan deviation is as low as $1.7\times10^{-13}$, a record result for a vapor-cell clock. As expected, the clock signal is affected by white frequency noise and consequently the Allan deviation scales as $\tau^{-1/2}$, $\tau$ being the integration time, and reaches the value of $6\times 10^{-15}$ after $\tau \approx 2000$ s. A drift of $-8\times10^{-15}/$day is removed from the data. For our cell the He permeation effect, that has a time constant of about 200 days, is considered extinguished and we attribute the frequency drift mainly to temperature. In particular, the observed value corresponds to a temperature drift $\leq 100 \ \mu$K/day. 

\
To give a physical insight into the effects leading to this result, we start analyzing the factors limiting the short-term frequency stability of the POP clock.

\begin{figure}[!h]
\begin{center}
\includegraphics[height=250pt]{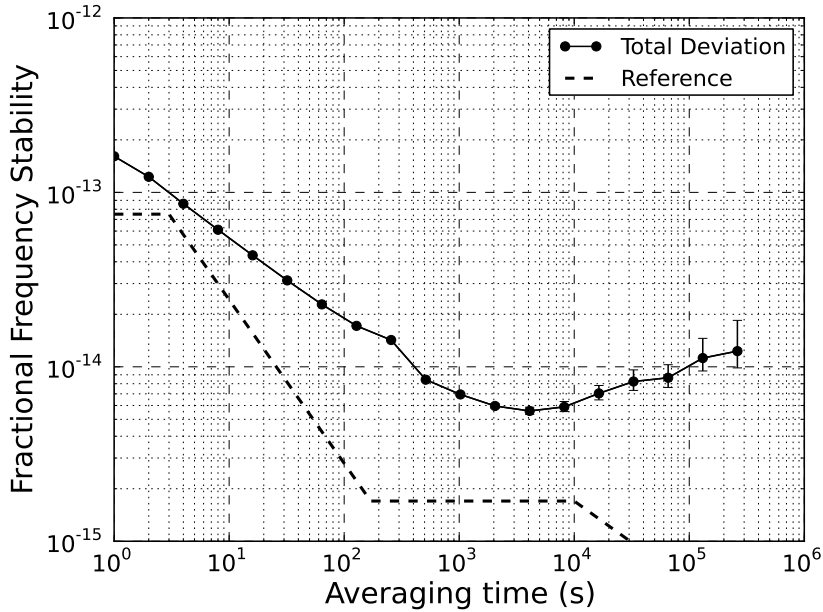}
\caption{POP clock frequency stability: typical medium-term measurement.}\label{}
\end{center}
\end{figure}
\
First, we consider the shot-noise associated to the detection of optical photons that provides the ultimate theoretical stability limit. In terms of Allan deviation, this noise contribution can be expressed as:

\begin{equation}
\sigma_{y}^{sn}(\tau)=\frac{1}{\pi \ Q_{a} \ R_{sn}}\sqrt{\frac{T_{C}}{\tau}}
\end{equation}

where $Q_{a}$ is the quality factor of the atomic resonance ($Q_{a}= 4.4\times10^{7}$) and $R_{sn}$ is the signal-to-noise ratio that is proportional to the contrast $C$ of the central Ramsey fringe. In our operating conditions, it turns out $R_{sn} \approx 20000$ and $\sigma_{y}^{sn}(\tau)= 2\times10^{-14} \ \tau^{-1/2}$.

\
We evaluated also the noise contribution due by the transimpedance amplifier (ti) that introduces an additive white frequency noise. It was measured in absence of laser light and gives a contribution of $\sigma_{y}^{ti}(\tau)=4\times10^{-14} \ \tau^{-1/2}$.

\
In frequency standards working in pulsed operation a well known issue is the phase noise of the microwave interrogating signal that is transferred to the atomic signal through the Dick effect \cite{santarelli1998}.

\begin{equation}\label{dick_laser}
\sigma_{y}^{LO}(\tau)=\left\{\sum_{k=1}^{\infty}\textnormal{sinc}^{2}\bigg(k
\pi\frac{T}{T_{C}}\bigg)S_{y}^{LO}(k f_{C})\right\}^{1/2}\tau^{-1/2}
\end{equation}

where $S_{y}^{LO}(f)$ is the power spectral density of the microwave fractional frequency fluctuations and $f_{C}=1/T_{C}$ ($f_{C}\approx 200$ Hz in our case). With the performances reported in subsection 4.1, the Dick effect results $\sigma_{y}^{LO}(\tau)= 7\times10^{-14} \ \tau^{-1/2}$.

\
Another effect limiting the short-term stability of the POP clock in the optical detection mode is the amplitude fluctuations of the laser probe through which the clock signal is observed (see also \cite{camparo1998, camparo1999}). Since this additive noise is sampled only during the detection time, its contribution to the clock Allan deviation can be written in a form similar to Dick effect as:

\begin{equation}\label{dick_laser}
\sigma_{y}^{AM}(\tau)=\frac{1}{C Q_{a}}\left\{\sum_{k=1}^{\infty}\textnormal{sinc}^{2}\bigg(k
\pi\frac{T}{T_{C}}\bigg)S_{AM}(k f_{C})\right\}^{1/2}\tau^{-1/2}
\end{equation}

where $S_{AM}(f)$ is the power spectral density of the fractional intensity fluctuations of the probe signal reaching the photodetector. It contains both the laser relative intensity noise (RIN) transferred at the output of the cell (AM-AM) and the laser frequency noise converted into amplitude fluctuations (PM-AM). We have measured 
$S_{AM}(f) = 2 \times 10^{-11} \ \textnormal{Hz}^{-1}$ for $100 \ \textnormal{Hz} < f< 1 $ kHz which reveals an excess of laser frequency noise converted into AM fluctuations at the cell output. Equation (10) gives then the following value for the Allan deviation: $\sigma_{y}^{AM}(\tau)=1.2 \times 10^{-13} \ \tau^{-1/2}$. 

\
It is worth mentioning that, since this is an additive noise, it scales as the contrast; this means that the larger is the contrast the lower is this noise contribution.

\
From the previous evaluation it turns out that the sum of all the previous contributions leads to $1.6\times10^{-13}$ at 1 s, very close to the measured value; moreover, the short-term clock stability is mainly limited by the laser noise.
 
\
Regarding the medium-long term performances, we think that the observed behavior after 2000 s of integration times is mainly due to thermal fluctuations, probably related to thermal bridges between the external environment and the physics package. To show that the clock signal is not affected by any flicker noise coming from the laser and/or the electronics we refer to the plot of Fig. 15.  This graph is obtained selecting a subset of 50000 points from the same run of Fig. 14. The plot shows that in quiet environmental conditions the clock stability may exhibit values below the $10^{-15}$  level. We are then confident that with a better engineering of the system, based on a more refined thermal model, the stability of Fig. 15 can become a typical performance.
%(The stability results previously mentioned place this clock among the best ever realized in the field of vapor-cell frequency standards, either using the optical pumping technique or the isotropic cooling, and it is not far from %the active H-maser performances.)
\begin{figure}[!h]
\begin{center}
\includegraphics[height=250pt]{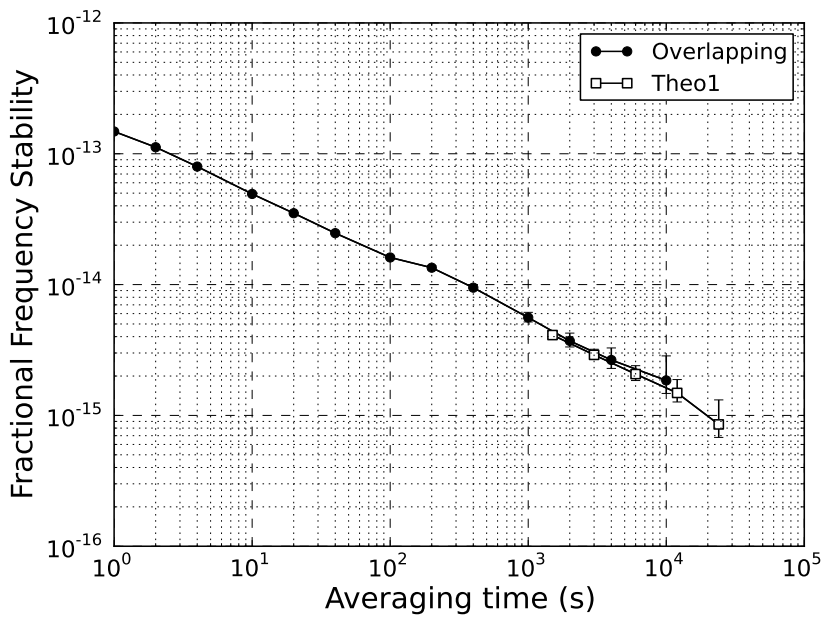}
\caption{POP clock frequency stability in quiet environmental conditions.}\label{}
\end{center}
\end{figure}

%(significant improvement in the ultimate stability limit signal-to-noise ratio higher than that achievable by microwave detection that is thermal noise limited. that sets with respect the ultimate physical stability value %achievable by the clock turns out one order of magnitude lower than thermal noise that sets carry more energy than then  it is expected that a detection in the optical occurs with a higher signal-to-noise ratio and consequently a %better short-term frequency stability.  )
%It is worth mentioning that the frequency stability reported in Fig. 15 has been observed several times and represents a typical clock performance.

\section*{8. Conclusions}

We have reported in this paper the implementation and the metrological characterization of a pulsed optically pumped Rb clock based on the optical detection mode. The design of the clock followed the guidelines of the theory developed in \cite{micalizio2009}. That theory, already successfully tested for the maser approach, proved its validity also in this case: physical behavior and frequency stability performances are correctly predicted. 

\
Specifically, the short-term stability currently turns out limited by the laser noise transferred to the amplitude of the clock signal. At first, this noise contribution may be lowered by a factor of two by using a better laser power supply. In addition, since this noise scales as the contrast, it is possible to adopt the total pumping technique in order to increase the number of pumped atoms and then the contrast \cite{micalizio2009b}, according to Eq. (10). With these improvements it is reasonable to reduce $\sigma_{y}^{AM}$ to the level of $3\times10^{-14}$ at 1 s.

\
The other main instability source comes from the phase noise of the interrogating microwave. Thanks to the spectral characteristics of ultra low-noise 100 MHz quartz oscillators and adjusting properly the timing sequence, the Dick effect contribution can be reduced to $3\times10^{-14}$ at 1 s. Definitely, the overall clock frequency stability can achieve the value of $5\times10^{-14}$ at 1 s.

\
As regards integration times in the medium- long-term period, the performances appear to be currently limited by the instability and non-homogeneity of the cell temperature. However, once solved this issue, we are confident that the clock can reach, at least in laboratory conditions, the region of $10^{-16}$, with a drift below $5\times10^{-15}/\textnormal{day}$.

\
Indeed, limiting ourselves to consider the results reported in the present paper, we point out that they already represent a record achievement for a vapor a cell clock.

\
To make more evident the importance of the result we attained, we show in the following figure a comparison between the POP clock frequency stability and the ground tests of the GALILEO's clocks \cite{waller2010}.
\begin{figure}[!h]
\begin{center}
\includegraphics[height=250pt]{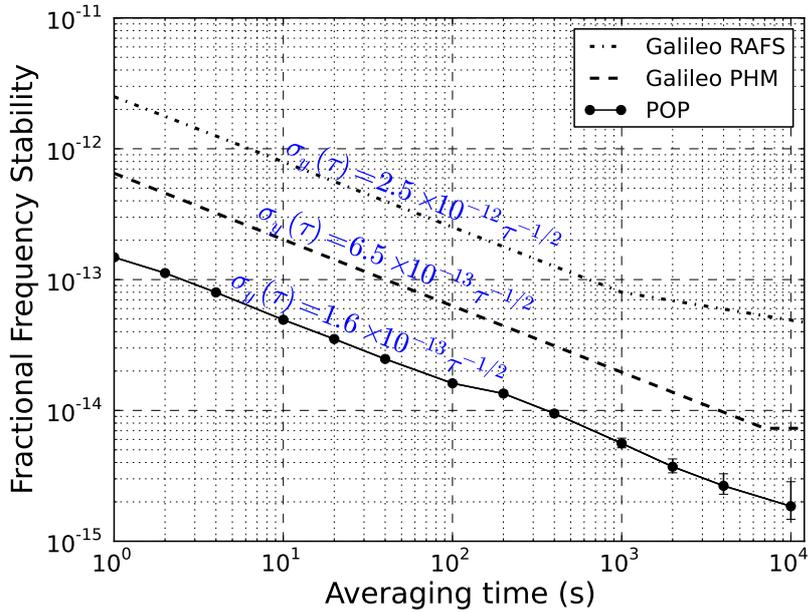}
\caption{Comparison between the POP clock stability and the performances of the Rubidium Atomic Frequency Standard (RAFS) and the PHM of the GALIELO system (ground tests).}\label{}
\end{center}
\end{figure}
\
It is observed that our clock exhibits a frequency stability even better than the performance of the PHM, with a physics package that can be more compact and lighter.

\
In addition, the result of Fig. 15 turns out particularly remarkable if we think that it is achieved with a hot atomic sample. A similar short-term stability ($2.2\times 10^{-13}$ at 1 s using a cryogenic oscillator as LO) is reported in \cite{esnault2010} where a cold-atom vapor-cell clock is described. However, that value is very close to the ultimate stability limit given by the atomic shot noise, while in our case the shot noise limit is still far from being reached, as just discussed.

\
Definitely, the POP clock combines typical features of vapor-cell devices, such as compactness, reliability, low power consumption, etc. with the high frequency stability performances of H-masers. All these features make the POP clock with optical detection very attractive for many technological applications and particularly suitable for a space-oriented engineering.

\section*{Acknowledgments}

This work has been funded by the European Space Agency(ESA contract "Next Generation Compact Atomic Clocks" 21504/08/NL/GLC).

\


\begin{thebibliography}{200}
\bibitem{esnault2010} Esnault F-X, Holleville D, Rossetto N, Guerandel S and Dimarcq N 2010 \textit{Phys. Rev. A} \textbf{82} 033436
\bibitem{micalizio2009} Micalizio S, Godone A., Levi F and Calosso C 2009 \textit{Phys. Rev. A} \textbf{79} 013403
\bibitem{bandi2011} Bandi T, Affolderbach C, Calosso C E, Mileti G 2011 \textit{Electronics Letters} \textbf{47} 698-699
\bibitem{godone2006}	Godone A,  Micalizio S, Levi F, Calosso C 2006 \textit{Phys. Rev. A} \textbf{74} 043401
\bibitem{mcguyer2009}	McGuyer B H, Jau Y-Y and Happer W 2009 \textit{Appl. Phys. Lett.}  \textbf{94} 251110
\bibitem{affolderbach2005} Affolderbach C, Andreeva C, Cartaleva S, Karaulanov T, Mileti G and Slavov D 2005 \textit{Appl. Phys. B} \textbf{80} 841-848
\bibitem{shah2006} Shah V, Gerginov V, Schwindt P D D, Knappe S, Hollberg L and Kitching J 2006 \textit{Appl. Phys. Lett.} \textbf{89} 151124
\bibitem{zhu2000} Zhu M and Cutler L 2000 \textit{Proceedings of the Precise Time and Time Interval (PTTI) Systems and Applications Meeting}, Reston, Virginia, USA, pp. 311-322.
\bibitem{zhu2003} Zhu M 2003 \textit{Proceedings of the Joint Meeting 17th European Frequency and Time Forum and 2003 IEEE International Frequency Control Symposium}, Edited by the IEEE Ultrasonics, Ferroelectrics, and Frequency Control Society, Tampa, Florida, 16-21.
\bibitem{godone2004} Godone A, Levi F, Micalizio S and Calosso C 2004 \textit{Phys. Rev. A} \textbf{70} 012508
\bibitem{nagel1999} Nagel A, Brandt S, Meschede D and Wynands S 1999 \textit{Europhys. Lett.} \textbf{48} 385
\bibitem{godone2007} Godone A, Levi F, Micalizio S and Calosso C 2007 \textit{IEEE Trans. Instrum. Meas.} \textbf{56} 378-382 
\bibitem{english1978}  English TC, Jechart E and  Kwon TM 1978 in
\textit{Proceedings 10th Precise Time and Time Interval Forum}, Washington, edited by LJ Rueger, (1979, NASA, Maryland) 147-65
\bibitem{alekseyev1974} Alekseyev E I, Bazarov Y N and Levishin E I 1974 \textit{Radio Eng. Electron. Phys.} \textbf{19} 77-83
\bibitem{alekseyev1975} Alekseyev E I, Bazarov Y N and Telegin G I 1975 \textit{Radio Eng. Electron. Phys.} \textbf{20} 73-80
\bibitem{vanier_audoin1985} Vanier J and Audoin C 1989 \textit{The Quantum Physics of Atomic Frequency Standards} (Bristol-England: Adam-Hilger)
\bibitem{vanier1982} Vanier J, Kunski R, Cyr N, Savard J Y and Tetu M 1982 \textit{J. Appl. Phys.} \textbf{53} 5387
\bibitem{godone2011} Godone A, Micalizio S, Levi F and Calosso C, 2011 \textit{Rev. Sci. Instrum.} \textbf{82}, 074703.
\bibitem{micalizio2010} Micalizio S, Godone A, Levi F and Calosso C 2010 \textit{IEEE Trans. Ultrason. Ferroelct. Freq. Control} \textbf{57} 1524
\bibitem{micalizioFCSEFTF} Micalizio S, Godone A, Calosso C, Levi F, Affolderbach C and Gruet F  \textit{IEEE Trans. Ultrason. Ferroelct. Freq. Control} accepted for publication
\bibitem{calosso2007} Calosso C, Micalizio S, Godone A, Bertacco EK and Levi F 2007 \textit{IEEE Trans. Ultrason. Ferroelct. Freq. Control} \textbf{54}, 1731-40
\bibitem{boudot2009} Boudot R, Guerandel S, De Clercq E, Dimarcq N and Clairon A 2009 \textit{IEEE Trans. Instrum. Meas.} \textbf{58} 1217 
\bibitem{santarelli1998} Santarelli G, Audoin C, Makdissi A, Laurent P, Dick G J and Clairon A 1998 \textit{IEEE Trans. Ultrason. Ferroelct. Freq. Control} \textbf{45} 887-89
\bibitem{camparo1998} Camparo J C 1998 \textit{J. Opt. Soc. Am. B} \textbf{15} 1177
\bibitem{camparo1999} Camparo J C and Coffer J G 1999 \textit{Phys. Rev. A} \textbf{59} 728
\bibitem{micalizio2009b} Micalizio S, Godone A, Levi F and Calosso C 2009 \textit{Phys. Rev. A} \textbf{80} 023419
\bibitem{waller2010} Waller P, Gonzalez S, Binda S, Sesia I, Hidalgo I, Tobias G and Tavella P 2010 \textit{IEEE Trans. Ultrason. Ferroelct. Freq. Control} \textbf{57} 738




\end{thebibliography}
\end{document}